\documentclass[a4paper,11pt]{article}

\usepackage[english]{babel}

\usepackage{authblk}
\usepackage{amsfonts}
\usepackage{amsmath}
\usepackage{graphicx}
\usepackage{caption}
\usepackage{subcaption}
\usepackage{amsthm}
\usepackage{bbm}
\usepackage{float}

\newtheorem{definition}{Definition}[section]

\newtheorem{proposition}{Proposition}[section]
\newtheorem{remark}{Remark}[section]

\newtheorem*{theorem*}{Theorem}
\newtheorem*{proposition*}{Proposition}

\newcommand{\PA}{P^{(a)}}
\newcommand{\PB}{P^{(b)}}
\newcommand{\TA}{T^{(a)}}
\newcommand{\TB}{T^{(b)}}
\newcommand{\CA}{C^{(a)}}
\newcommand{\CB}{C^{(b)}}
\newcommand{\LA}{L^{(a)}}
\newcommand{\LB}{L^{(b)}}

\newcommand{\E}{\mathbb{E}}
\newcommand{\I}{\mathbb{I}}

\title{Estimation of slowly decreasing Hawkes kernels: \\Application to high frequency order book modelling}

\author[1]{Emmanuel Bacry}
\author[1]{Thibault Jaisson}
\author[1,2]{Jean-Fran\c{c}ois Muzy}
\affil[1]{\small Centre de Math\'ematiques Appliqu\'ees, CNRS, \'Ecole Polytechnique, \authorcr UMR 7641, 91128 Palaiseau, France}
\affil[2]{\small Laboratoire Sciences Pour l'Environnement, CNRS, Universit\'e de Corse, \authorcr UMR 6134, 20250 Cort\'e, France}
\date{}  

\begin{document}

\maketitle

\begin{abstract}
\noindent We present a modified version of the non parametric Hawkes kernel estimation procedure studied in \cite{bacry2014second} that is adapted to slowly decreasing kernels. We show on numerical simulations involving
a reasonable number of events that this method allows us to estimate
faithfully a power-law decreasing kernel over at least 6 decades.
We then propose a 8-dimensional Hawkes model for all events associated with the first level of some asset order book. Applying our estimation procedure to this model, allows us to uncover the main properties of the coupled
dynamics of trade, limit and cancel orders in relationship with the mid-price variations.
 \end{abstract}

\noindent \textbf{Keywords:} Hawkes processes, kernel estimations, power law kernels, high frequency, order book dynamics, trades, limit orders, cancel orders, market impact.

\section{Introduction}

The understanding of the price formation mechanism remains one of the most challenging problem in quantitative finance. In most modern financial exchanges, assets are traded via a continuous double auction: agents can choose to buy or sell a stock at certain prices posting \textit{limit orders} in the order book or to execute the available limit orders using \textit{market orders}. As long as their their limit orders are not executed, agents can also use \textit{cancel orders} to take them off\footnote{See e.g., \cite{cont2013price,farmer2003doubleauction} for more details on the functioning of continuous double auctions.}.
The intrinsic complexity of the formation of order book (it is the result of market, limit and cancel orders of a large number of anonymous traders) implies that in spite of its great practical and theoretical importance, there are relatively few full order book models (see \cite{rocsu2009dynamic,farmer2003doubleauction} for ``zero-intelligence'', i.e. purely random model of book dynamics and \cite{cont2013markov,gareche2013fokker,huang2013simulating} for Markovian models of the order book).\\

\noindent A class of simpler models does not take into account the amount of liquidity and focuses exclusively on asset prices which are described as the result
of the {\em impact} of various type of orders: Former models considered exclusively the impact of market orders \cite{bouchaud2009markets,bouchaud2004fluctuations,jaisson2014market} while some recent models account for the influence of all types of order book events \cite{cont2013price,eisler2012price}. In a recent work, some of us \cite{bacry2014hawkes} introduced a price impact model using the framework of Hawkes processes. This model shares features with former impact models but allows one to describe the joint dynamics of both
trades and price moves. In particular, unlike previous impact models, 
this Hawkes model takes into account the influence of price changes on themselves and on trades.\\

\noindent Hawkes processes were defined in \cite{hawkes1971point} as point processes whose intensity function is a linear regression on the past of the process
$$\lambda^N_t=\mu+\int_{(-\infty,t)} \phi(t-s)dN_s.$$
The self and mutually exciting nature of Hawkes processes makes them naturally adapted to the modelling of multivariate counting processes. They have for example been successfully applied in many domains such as the study of earthquakes \cite{ogata1999seismicity}, neurobiology \cite{reynaud2014goodness} or sociology \cite{mohler2011self}.
In Finance, they have been recently used to model the arrival of trades or limit orders \cite{hewlett2006clustering,bowsher2007modelling,large2007measuring}, price variations at microstructure level \cite{bacry2013modelling,bacry2014hawkes,hardiman2013critical}, financial contagion \cite{ait2010modeling} or credit risk \cite{errais2010affine}.\\

\noindent In this work, we will propose a generalization of the impact model of \cite{bacry2014hawkes}, in order to account for the interaction between price variations and any event (market, limit or cancel order)
at the level I of the order-book. This model turns out to be
an 8-dimensional Hawkes process that will allow us to have a direct measure of the endogeneity and the causality between these events, see \cite{filimonov2012quantifying} or \cite{hardiman2013critical} for related discussions about the use of Hawkes processes to measure the market endogeneity.\\

\noindent As far as estimation problems are concerned, the few former non parametric estimates of Hawkes kernels in the context of high frequency finance lead to power law decreasing kernels whose exponent is slightly higher than one \cite{bacry2012non,bacry2014hawkes,hardiman2013critical}. As we will see, this implies that one needs to estimate the kernels on a wide range of time scales (typically from 10 microseconds to 100 seconds) to capture the dynamics of the order book. For that purpose we propose an improvement of the numerical scheme introduced in \cite{bacry2014hawkes,bacry2014second} in order to handle slowly decreasing kernels. It will be illustrated on some numerical examples, that this method allows us to estimate the kernel on a large range of time scales which cannot be done within the Gaussian scheme of \cite{bacry2014second}.\\


\noindent The paper is organized as follows. After recalling the main definitions and properties of multidimensional Hawkes processes, we present the main non-parametric estimation principles in Section \ref{s1}. Section \ref{s2} explains how these principles can be adapted to the non-parametric estimation of slowly decaying kernels. The so-obtained algorithm is presented and tested on numerical simulations at the end of this Section. The level I book model is then introduced in Section \ref{s3}. It is calibrated on high frequency data corresponding to DAX and BUND front future contracts during one year (from June 2013 to June 2014). We then comment our results as far as endogeneity and mutual influence of all types of events are concerned. Discussion in relationship with former works and prospects for future research are provided in the concluding Section \ref{conc}.

\section{Multidimensional Hawkes processes and estimation principles}
\label{s1}
In this section we recall the definition of a $D$-dimensional Hawkes processes
and briefly review its main properties. For more details we refer
the reader to \cite{bacry2014second}.

\subsection{Basic definitions}

\begin{definition}
A D-dimensional point process $N_t=(N^i_t)_{1\leq i\leq D}$ is a Hawkes process if for every $i$, the
$i^{th}$ component of its
intensity function $\lambda_t$ is a linear regression of the past jumps of $N_t$, i.e.,
\begin{equation}
\label{defHawkes}
\lambda^i_t=\mu^i+\sum_{j=1}^D\int_{(-\infty,t)} \phi^{ij}(t-s)dN_s^j, ~~~\forall i \in [1,D],
\end{equation}
where $\mu=\{\mu^i\}_{1\leq i\leq D}$ is the so-called
{\em exogenous intensity} (with $\mu^i \geq 0$), and
$\phi=\{\phi^{ij}\}_{1\leq i,j\leq D}$ the so-called {\em Hawkes kernel} matrix, where
each $\phi^{ij}(t)$ is a real positive and causal function\footnote{By definition, a function is causal if its support is included in $\mathbbm{R}^+$.}.
Using matrix convolution notation, the system of Equation \eqref{defHawkes} simply rewrites
\begin{equation}
\label{defHawkes1}
\lambda_t=\mu+\phi \ast dN_s.
\end{equation}
\end{definition}

\noindent Let us recall the following well known stability condition of the Hawkes processes, see for example \cite{hawkes1971spectra}.
\begin{proposition}
\label{prop1}
If the matrix of the norms $L^1$ of the kernels defined as $$||\phi||_=(||\phi^{ij}||_1)_{1\leq i,j\leq D}$$ has a spectral radius strictly smaller than one then $N$ admits a version with stationary increments.
\end{proposition}

\noindent Let us remark that Br\'emaud and Massouli\'e \cite{bremaud1996stability} have shown that this stability
criterion remains valid when, in Equation \eqref{defHawkes}, $\lambda_t$ is a nonlinear positive Lipschitz function of $\{dN_s\}_{s<t}$ and where the elements of the kernel $\phi$ are not restricted to be positive. A particularly interesting generalization of Hawkes processes considered in \cite{hansen2012lasso} is
\begin{equation}
\label{defHawkes2}
\lambda_t= \left(\mu+\phi\ast dN_s \right)^+
\end{equation}
where $(x)^+ = x$ if $x>0$ and $(x)_+ = 0$ otherwise.
This extension allows one to account for inhibitory effects when $\phi^{ij}(t) < 0$.\\

\noindent We will now assume that the condition of Proposition \ref{prop1} is satisfied and consider a stationary Hawkes process.
The next simple result describes the behavior of the first order statistics of $N$, see for example \cite{bacry2014second}.

\begin{proposition}
\label{order1}
The average of the intensity vector of $N$: $$\Lambda=(\E[\lambda^1_t],...,\E[\lambda^D_t])^T$$ satisfies
\begin{equation}
\label{mu}
\Lambda=(\mathbb{I}-||\phi||)^{-1}\mu.
\end{equation}
\end{proposition}

\subsection{Second-order properties and Wiener-Hopf system}
\label{sec:secondorder}
The second order statistics of Hawkes processes are naturally characterized by their infinitesimal covariance functions.
\begin{definition}
The infinitesimal covariance $\nu$ of $N$ is the $D \times D$ matrix whose elements are
$$\nu^{ij}(t-t')dtdt'=\E[dN^i_tdN^j_{t'}]-\Lambda^i\Lambda^jdtdt',$$
where $\Sigma$ is the $D \times D$ matrix whose diagonal is the average intensities $\Lambda^i$.
\end{definition}
\noindent The following result is shown in \cite{bacry2012non}.
\begin{proposition}
$\nu$ can be expressed as a functional of $\phi$:
\begin{equation}
\label{nuphi}
\nu = (\delta \mathbb{I} + \psi) \ast \Sigma (\delta \mathbb{I} + \psi^T)
\end{equation}
where $\delta(t)$ stands for the Dirac distribution, $\Sigma$ is the $D \times D$ matrix whose diagonal is the average intensities $\Lambda^i$,
and
where the matrix $\psi(t)$ is
is defined as:
\begin{equation}
\label{defpsi}
\psi(t) = \sum_{k\geq 1} \phi^{(*k)}(t),
\end{equation}
in which we used the notation $\phi^{(*k)}(t) = \phi \ast \phi \ast \ldots \ast \phi$ (where $\phi$ is repeated $k$ times).
\end{proposition}

\noindent Let us mention that there is an alternative ``population representation'' of Hawkes processes (see e.g., \cite{hawkes1974cluster}) according to which
they are built as the times of arrival and birth of the following population process:
\begin{itemize}
\item{There are individuals of type $1$, ..., $D$.}
\item{Migrants of type $i$ arrive at a Poisson rate of $\mu^i$.}
\item{Every individual can have children of (a priori) all types and the children of type $i$ of an individual of type $j$ who was born or migrated in $t$ are distributed as an inhomogeneous Poisson process of intensity $\phi^{ij}(\cdot -t)$. }
\end{itemize}
In this model, $||\phi^{ij}||$ appears as the average number of children of type $i$ of an individual of type $j$. We thus use $||\phi^{ij}||$ as a measure of ``causality'' between $j$ and $i$. Similarly, $\Lambda^j/\Lambda^i ||\phi^{ij}||$ appears as the proportion of individuals of type $i$ whose parent is an individual of type $j$.\\

\noindent In this construction, the elements $\psi^{ij}$ of the matrix $\psi$ also have a natural interpretation. The descendants of type $i$ of each exogenous event of type $j$ occurring at time $t$ are a point process of average intensity $\psi^{ij}(\cdot-t)$. Therefore, $||\phi^{ij}||$ appears as the average number of individuals of type $i$ which descend from an individual of type $j$. Similarly, $\Lambda^j/\Lambda^i ||\psi^{ij}||$ appears as the proportion of individuals of type $i$ whose ``ancestor'' is an individual of type $j$.\\

\noindent As explained in \cite{bacry2014second}, since the jumps are always of size one, the second order statistics of the Hawkes process can also be summed up by its conditional laws.
\begin{definition}
\label{def_claw}
We define the conditional laws denoted $(g^{ij})_{1\leq i, j\leq D}$ as the non singular part of the measure $\E[dN^i_{t}|dN^j_0=1]$:
$$g^{ij}(t)dt=\E[dN^i_{t}|dN^j_0=1]-\mathbbm{1}_{i=j}\delta(t)-\Lambda^idt.$$
They are linked with the infinitesimal covariance by the equation
\begin{equation}
\label{nug}
g(t) = \nu(t) \Sigma^{-1} -\delta(t)\I.
\end{equation}
\end{definition}
\noindent The following proposition, see \cite{bacry2014second}, links the conditional laws and the kernels of Hawkes processes through a Wiener-Hopf system of equations.
\begin{proposition}
Given the conditional laws of a Hawkes process $g$, its kernel $\phi$ is the only\footnote{The uniqueness of the solution of Equation \eqref{wh}, considered as an equation in $\phi$, is shown in \cite{bacry2014second}.} solution of the Wiener-Hopf system:
\begin{equation}
\label{wh}
g(t)=\phi(t)+g\ast\phi(t), ~~\forall t>0.
\end{equation}
\end{proposition}

\begin{remark}
The linear structure of Hawkes processes implies that they have properties similar to that of autoregressive processes. In particular, the previous proposition can be seen as the counter part of the Yule-Walker Equation.
\end{remark}

\subsection{Existence and uniqueness of solutions of the Wiener-Hopf system}
\label{existence}
The Wiener-Hopf system \eqref{wh} is at the core of our estimation procedure.
In \cite{bacry2014second}, it is shown that if $g$ is the conditional law of a Hawkes process as defined in Definition \ref{def_claw} then the system \eqref{wh} admits the unique causal solution $\phi$. \\


\noindent Let us now assume that $\Sigma$, $\nu$ and $g$ are respectively the average, the infinitesimal covariance function and the conditional law of any stationary $D$-dimensional point process (not necessarily a Hawkes process). Using the Wiener-Khintchine Theorem, we get that the matrix $\widehat{\nu}(z)$ is positive definite for any $z\in \mathbb{R}$. We can thus apply Theorem 8.2 of \cite{gokhberg1958systems} to decompose $\nu$ as $$\nu(t)=Y(t)\ast Y^T(-t)$$
where $Y$ is causal. We then set $$\psi=(Y-\delta \mathbb{I})\sqrt{\Sigma^{-1}}$$
and $\phi$ such that
$$\hat{\phi}=\hat{\psi}(\mathbb{I}+\hat{\psi})^{-1}$$
to get that $\phi$ satisfies Equation \eqref{nuphi} and thus Equation \eqref{wh} (these two equations are equivalent for $g(t) = \nu(t) \Sigma^{-1} -\delta(t)\I$).\\

\noindent Therefore, even if the point process that we are studying is not a Hawkes process Equation \eqref{wh} has one and only one solution. It thus makes sense to try to find the solution of the Wiener-Hopf system for empirical data\footnote{Even if the empirical conditional laws are not equal to the real conditional laws.}.
Of course, the so-obtained solution is no longer necessarily positive. We will comment the interpretation of a negative $\phi$ later.

\begin{remark}
It is noteworthy that, from a fundamental point of view, solving the Wiener-Hopf system amounts to finding the kernel matrix $\phi$ 
(and the intensity vector $\mu$) that provide the best linear predictor (in the sense of mean-square error) 
of the conditional intensity. This remark leads the authors of \cite{hansen2012lasso,reynaud2014goodness} to consider an alternative estimation algorithm relying on the minimization
of a contrast function, a proxy for the mean square error (see \cite{bacry2014second} for more details). 
\end{remark}

\subsection{Estimation principles}
\label{principles}
As explained in \cite{bacry2014second}, the procedure for solving numerically the Wiener-Hopf system \eqref{wh} follows Nystr{\"o}m method \cite{nystrom1930praktische}. More precisely, given
a realization  if the process $N$ on an interval $[0,T]$, the procedure for
non parametric estimation is the following
\begin{itemize}
\item[1.] Compute an estimation $\tilde g$ of the matrix function $g$ using a ``fine enough'' time grid (values of $g$ outside of this time grid will be computed using an interpolation scheme).

\item[2.] Use a quadrature method to discretize the Wiener-Hopf system \eqref{wh}
on an interval $[T_{min},T_{max}]$. If we use the quadrature points $\{t_k\}_{1\le k \le K}$ along with the quadrature weights $\{w_k\}_{1\le k \le K}$, one gets, the system
\begin{equation}
\label{nscheme}
\tilde g^{ij}(t_n) =  \tilde \phi^{ij}(t_n) + \sum_{l=1}^{D}\sum_{k=1}^{K} w_k \tilde g^{il}(t_n-t_k) \tilde  \phi^{lj}(t_k),~~~\forall n \in [0,K],
\end{equation}
for all $i,j \in [1,D]$.
\item[3.] Inverse this so-obtained $KD^2$ linear system. This leads to the estimation of the matrix kernel at the quadrature points
$\{\tilde \phi^{ij}(t_n)\}_{i,k\in[1,D],n \in [1,K]}$ as well as an estimation of $||\phi||_1$ (using quadrature).
\item[4.] Estimate empirically the average intensity $\Lambda$ (just counting the number of points on each component of the process) and estimate $\mu$ using \eqref{mu}.
\end{itemize}
A thorough study of this algorithm has been done in \cite{bacry2014second} (in the framework on a Gaussian quadrature) and  showed that it performs a fast and efficient non parametric estimation of a Hawkes process. It proved particularly competitive in the case the realization process has a large number of jumps and the matrix kernel is not well localized in time. Let us point out that \cite{bacry2014second} showed that this algorithm leads to precise result even if some elements of the kernel matrix have moderate negative values.\\

\noindent However, as we will show in the next section, when some elements of the kernel matrix are power-law, the quality of the estimation can be improved significantly. In this case, the use of Gaussian quadrature is not adapted and leads to inaccurate estimations because the ``mass'' of the power-law kernel exponents are spread over a large range of scales.
We shall study this phenomenon in the framework of high frequency financial data.

\section{Non parametric estimation of slowly decaying kernels}
\label{s2}
\label{plkernels}
One of the main issue when dealing with high frequency financial data is that the range of scales involved in the dynamics is rather large. This translates into kernels with large support,
typically from $T_{min}=100$ microseconds to $T_{max}=100$ seconds, slowly decaying (power-law like),
varying very quickly at small lags (around $T_{min}$) and much slower at larger lags (around $T_{max}$).
Let us follow step by step the algorithm of Section \ref{principles} and explain how to adapt it to this
framework.

\subsection{Using an adapted sample-grid for $g$}
\label{linearinterp}
Step 1. of the algorithm (estimation of $g$) is clearly the first sensitive step.
Since the kernels are varying very quickly at small lags, more care should be taken for estimating the behavior of $g$ in the neighbourhood of 0 than anywhere else.
In Appendix \ref{app_est}, we describe precisely the algorithm we shall use in this paper for the estimation of $g$. It basically consists in two step procedure: (i) empirical computation of $g$ on a time-grid which is more dense around 0 than anywhere else, (ii) linear interpolation for computing the values of $g$ at any time (in all our numerical experiments, we have checked than higher-order approximation does not bring relevant precision to the overall algorithm).


\subsection{Towards an adapted quadrature scheme}
The choice of the quadrature (step 2. in the algorithm) is the next (and last) sensitive step.
The Gaussian quadrature used in \cite{bacry2014second} for solving the Wiener-Hopf system is clearly not adapted to this situation. Indeed,
the Gaussian quadrature points are roughly uniform over $[0,T]$ whereas the kernels are varying very quickly around 0 and slowly anywhere else.
In order to get precise estimation of kernels on the whole range $[T_{min},T_{max}]$, the Gaussian quadrature is constrained by their behavior at small lags ($t \simeq T_{min}$) and would imply using an number of quadrature points of the order of $K=T_{max}/T_{min}=10^6$ which, even for $D=1$, is far to many to be able to solve numerically the system \eqref{nscheme}.\\

\noindent One way to circumvent this difficulty could consist in adding points close to zero by using the change of variable proposed in \cite{sloan1981quadrature}. Let $\{t_k,w_k\}_{1\le k \le K}$ correspond to the Gaussian quadrature scheme on $[\log(T_{min}),\log(T_{max})]$, we then set
\begin{equation}
\label{change}
(t'_k,w'_k)=(e^{t_k},e^{t_k}w_k).
\end{equation}
We then replace the system \eqref{nscheme} by the new
system:
\begin{equation}
\label{whcv}
\tilde g^{ij}(t'_n)=\tilde \phi^{ij}(t'_n)+\sum_{l=1}^D \sum_{k=1}^{K}w'_k\tilde g^{il}(t'_n-t'_k)
\tilde \phi^{lj}(t'_k).
\end{equation}
In order to test this new estimation scheme, we simulate
a one-dimensional Hawkes process with parameters
\begin{equation}
\label{hex}
\phi(t)=\frac{0.06}{(0.005+t)^{1.3}}~~~{\mbox{and}}~~~\mu = 0.05.
\end{equation}
The simulation is performed on a time-period long enough ($10^7$ seconds) so that the Monte Carlo error is small enough compared to the ``quadrature error''.

\begin{remark}
This kernel checks the same kind of ``multiscale behavior'' as kernels involved in high frequency financial dynamics. Indeed, 90\% of its $L^1$ norm is spread over 5 decades: from $10^{-3}$ to $10^2$ seconds, in the sense that $\int_{0}^{+\infty}\phi(t)dt=0.98$, $\int_{0}^{10^{-3}}\phi(t)dt=0.05$ and $\int_{10^2}^{+\infty}\phi(t)dt=0.05$.
\end{remark}

\begin{figure}[H]
\centering
        \begin{subfigure}[b]{0.49\textwidth}
                \includegraphics[width=\textwidth,height=50mm]{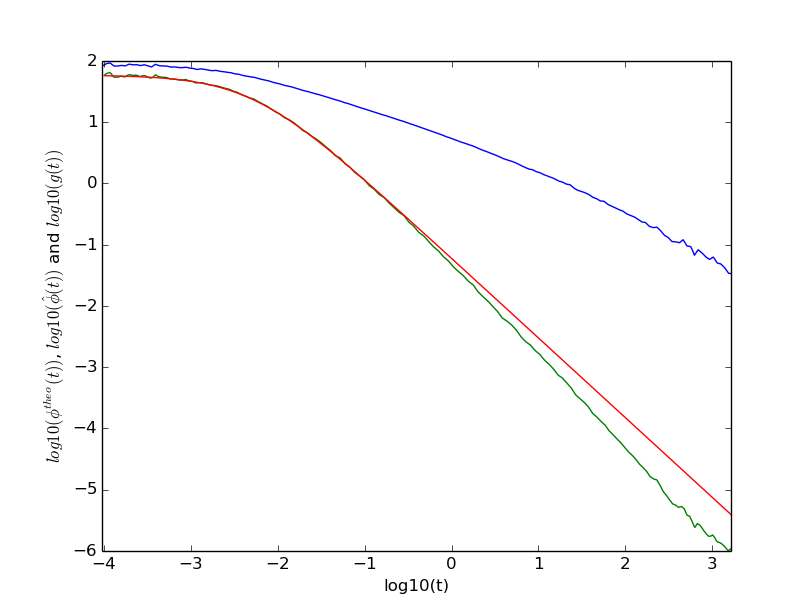}
        \end{subfigure}
        \begin{subfigure}[b]{0.49\textwidth}
                \includegraphics[width=\textwidth,height=50mm]{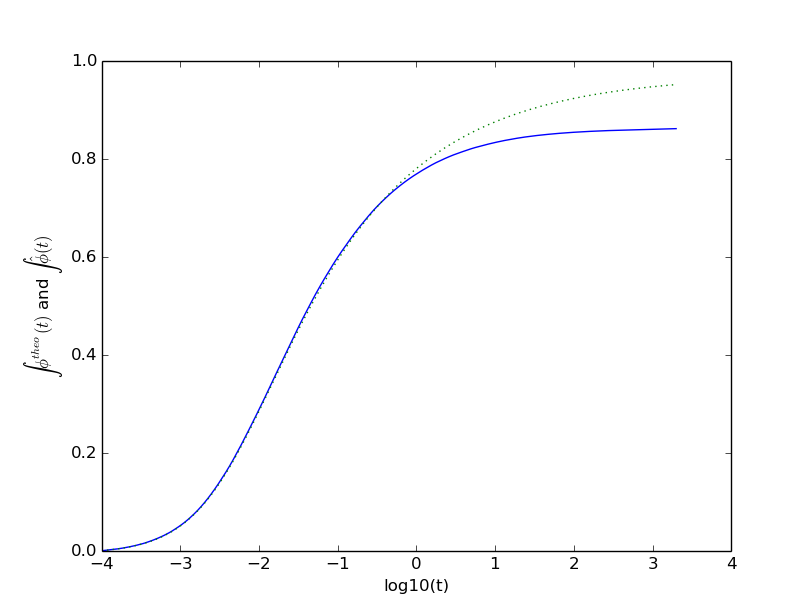}
        \end{subfigure}

        \caption{Left: Estimation of the conditional law $g$ and kernel $\phi$ in the case of a 1-dimensional Hawkes of size $T=10^7$ seconds, with parameters \eqref{hex}. The estimation of $g$ is performed using the procedure described in  Appendix \ref{app_est} using $h_{min}=1ms$, $h_{max}=1000s$ and $h_\delta=0.05$, and kernel estimation is performed using algorithm of Section \ref{nscheme} with Gaussian quadrature followed by the change of variable \eqref{change}
($K=200$, $T_{min}=1ms$, $T_{max}=2000s$ and thus $\delta=0.077$). The three curves correspond to
$\log_{10}-\log_{10}$ empirical conditional law  (blue), theoretical kernel (red) and estimated kernel (green) for 200 quadrature points.\\
Right: Cumulated theoretical kernel $\int_0^t\phi(s)ds$ (green dots) and cumulated estimated kernel $\int_0^t\tilde{\phi}(s)ds$ (blue) as a function of $\log_{10}(t)$.}
\label{llsim1}
\end{figure}

\noindent We applied the algorithm described in Section \ref{principles}, in which
we use the procedure described in Appendix \ref{app_est} to estimate $g$ using $h_{min}=1ms$, $h_{max}=1000s$ and $h_\delta=0.05$, and in which we replaced \eqref{nscheme} by \eqref{whcv}. We chose $K=200$, $T_{min}=1ms$ and $T_{max}=2000s$.
 Figure \ref{llsim1}, displays the so-obtained results. On the left, the log-log plot of both the estimated kernel and the theoretical kernel (as well as the conditional law $g$) are displayed. On the right, the integral on $[0,t]$ of both theoretical and estimated kernels are displayed.

%

\noindent We observe that, for large times ($t\geq 1s$), the estimated kernel does not approximate well the theoretical kernel. Moreover, this procedure significantly underestimates the cumulated kernel. Indeed, we get $||\tilde{\phi}||_1=0.85$ instead of $||\phi||_1=0.98$. This error on the estimation of the norms will be very important in our study of the causality between market events. Indeed the norms of the kernels $||\phi^{ij}||_1$ will have a direct interpretation as the average number of events of type $i$ caused by an event of type $j$.\\

\noindent Let us try to give an intuitive explanation of the bad performance of this method: Recall that the quadrature approximates
$$\int_{0}^{+\infty}g(t'_n-s)\phi(s)ds \simeq \sum_{k=1}^{K}w'_k g(t'_n-t'_k)\phi(t'_k), ~~~\forall n \in [1,K]$$
Thanks to the change of variable \eqref{change} there are more quadrature points around 0 than around any other time.
That ensures that the quadrature approximation captures the fast variation of $\phi(s)$ around $s=0$.
However, $g$ is also varying quickly around 0, the estimation error illustrated in Figure \ref{llsim1} comes from the bad approximation of $g(t'_n-s)$ around $s=t'_n$ (i.e., $k=n$ in the previous quadrature formula).\\

\noindent In the next section, we propose a new quadrature scheme that will capture all behaviour of the conditional law $g$ next to $t_n$ for any $n$.

\subsection{The adapted quadrature scheme for slow-decaying kernels}
Let us consider the time grid which is uniform from $0$ to $T_{min}$ and log-uniform between $T_{min}$ and $T_{max}$ using the same grid-size $\delta<1$
\begin{equation}
\label{points}
\{t_k\}_{1\le k\leq K}=[0,\delta T_{min},2\delta T_{min},...,T_{min},T_{min} e^{\delta},T_{min} e^{2\delta},...,T_{max}].
\end{equation}
The adapted quadrature scheme consists in considering that the kernels are piecewise affine on $[t_k,t_{k+1}]$:
$$\phi(t)=\phi(t_k)+\frac{t-t_k}{t_{k+1}-t_k}(\phi(t_{k+1})-\phi(t_k)).$$
Under this linear hypothesis, Equation \eqref{wh}, in the $D=1$ dimensional case, becomes (the generalization to dimension $D>1$ is straightforward):
\begin{eqnarray*}
g(t_n)&=&\phi(t_n)+\sum_{k=0}^{N-1} \phi(t_k) \int_{t_k}^{t_{k+1}}g(t_n-s)ds\\
&+&\sum_{k=0}^{N-1} (\phi(t_{k+1})-\phi(t_k)) \int_{t_k}^{t_{k+1}} \frac{s-t_k}{t_{k+1}-t_k} g(t_n-s)ds\\
&=&\phi(t_n)+\sum_{k=0}^{N-1} \phi(t_k) \int_{t_n-t_{k+1}}^{t_n-t_{k}}  g(u)du\\
&+&\sum_{k=0}^{N-1} \frac{\phi(t_{k+1})-\phi(t_k)}{t_{k+1}-t_k} \int_{t_n-t_{k+1}}^{t_n-t_{k}} (t_n-t_k-u) g(u)du
\end{eqnarray*}
Thus, using the adapted quadrature, the system \eqref{nscheme} becomes (in the $D=1$ case)
\begin{eqnarray}
\label{thequad}
\nonumber
\tilde g(t_n)&=&\tilde \phi(t_n)+\sum_{k=0}^{N-1} \tilde \phi(t_k) \int_{t_n-t_{k+1}}^{t_n-t_{k}}  \tilde  g(u)du\\
\nonumber
&+&\sum_{k=0}^{N-1} \frac{(\tilde \phi(t_{k+1})-\tilde \phi(t_k))(t_n-t_k)}{t_{k+1}-t_k} \int_{t_n-t_{k+1}}^{t_n-t_{k}}  \tilde  g(u)du\\
&-&\sum_{k=0}^{N-1} \frac{\tilde \phi(t_{k+1})-\tilde \phi(t_k)}{t_{k+1}-t_k} \int_{t_n-t_{k+1}}^{t_n-t_{k}} u \tilde g(u)du.
\end{eqnarray}
\noindent The empirical values of the integrals $\int_0^x \tilde g(u)du$ and $\int_0^x u \tilde g(u)du$ can easily be computed beforehand at all abscissa of the form $x=t_n-t_k$ using the linear interpolation of the conditional laws described in Section \ref{linearinterp} (see also Appendix \ref{app_est}).\\

\noindent As before, this is a linear equation in the values of $\phi$ at the quadrature points. Solving it, one gets an estimation of $\phi$ at these points. A simple linear interpolation can be performed to get estimation at other points.\\

\begin{figure}[H]
\centering
        \begin{subfigure}[b]{0.49\textwidth}
                \includegraphics[width=\textwidth,height=50mm]{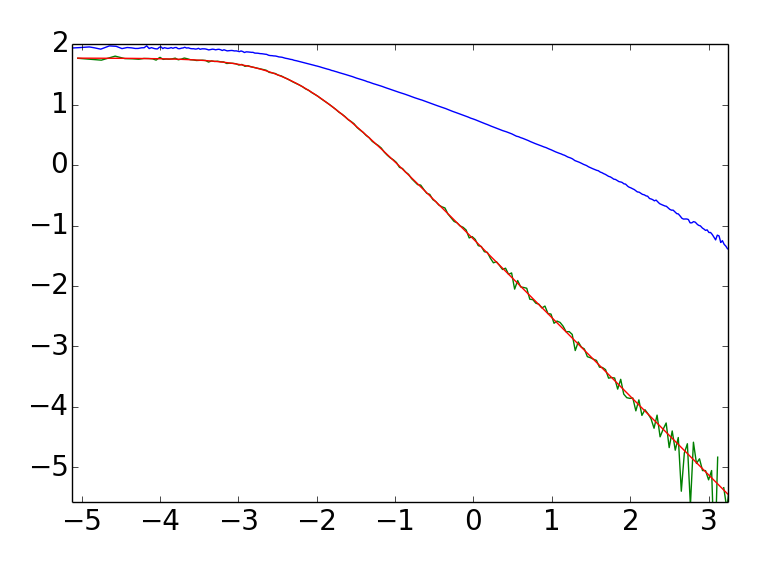}
        \end{subfigure}
        \begin{subfigure}[b]{0.49\textwidth}
                \includegraphics[width=\textwidth,height=50mm]{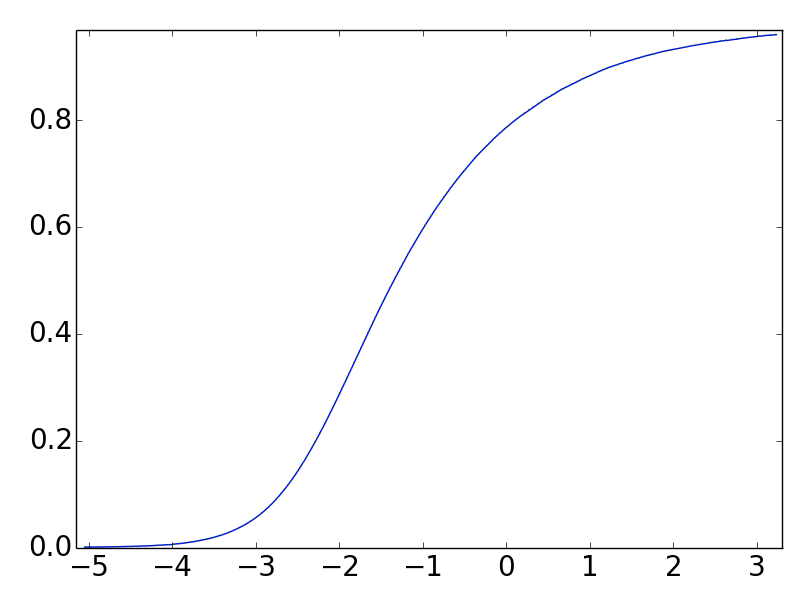}
        \end{subfigure}
        \caption{Left: Estimation of the conditional law $g$ and kernel $\phi$ for the same Hawkes process realization
as in Figure \ref{llsim1}.
The kernel estimation is performed using the algorithm of Section \ref{nscheme}
where the quadrature of step 3. has been replaced by
the adapted quadrature given by \eqref{thequad} ($K=200$, $T_{min}=1ms$, $T_{max}=2000s$ and thus $\delta=0.077$). The three curves correspond to
$\log_{10}-\log_{10}$ empirical conditional law  (blue), theoretical kernel (red) and estimated kernel (green) for 200 quadrature points. The estimation and the theoretical kernels perfectly match.\\
Right: Cumulated theoretical kernel $\int_0^t\phi(s)ds$ (green dots) and cumulated estimated kernel $\int_0^t\tilde{\phi}(s)ds$ (blue) as a function of $\log_{10}(t)$.}
\label{linscheme}
\end{figure}

\noindent Applying the same testing procedure as before, we get, see Figure \ref{linscheme} that this scheme works perfectly with a reasonable number of quadrature points.

%
%

\subsection{The adapted estimation procedure}
\label{thealgo}
Here is the step by step estimation procedure that will be used all along this paper:
\begin{itemize}
\item[1.] Compute an estimation $\tilde g$ of the matrix function $g$ using
the procedure described in Section \ref{linearinterp}.
\item[2.] Use the adapted quadrature method to discretize the Wiener-Hopf system \eqref{wh}
on an interval $[T_{min},T_{max}]$. The quadrature points are given by \eqref{points} and the scheme by \eqref{thequad} (this last equation corresponds to the one dimensional scheme, but generalization to any dimension is obvious).
\item[3.] Inverse this so-obtained $KD^2$ linear system. This leads to the estimation of the matrix kernel at the quadrature points
$\{\tilde \phi^{ij}(t_n)\}_{i,k\in[1,D],n \in [1,K]}$ as well as an estimation of $||\phi||_1$ (using quadrature).
\item[4.] Estimate empirically the average intensity $\Lambda$ and estimate $\mu$ using \eqref{mu}.
\end{itemize}

\section{Hawkes model for Level I Order book events}
\label{s3}

\subsection{Definition of the model}
\label{obmodel}
Multivariate Hawkes models provide a natural framework
to account for the impact of past events of various type on the rate of arrival of future events.
In \cite{bacry2014hawkes}, this framework has been considered to extent classical impact price
models (as in e.g. \cite{bouchaud2004fluctuations}) by accounting in real time for
the ``impact'' of market orders on price changes but also for the auto-regressive dynamics
of price changes and their retro-action on the rate of market order arrivals.
Along the same line, one can extend the model of \cite{bacry2014hawkes} by accounting, within
a multidimensional Hawkes model, for the cross and self influencing dynamics
of all event types in the order book. For the sake of parsimony and simplicity, we consider
only events occurring at the Level I of the order book, i.e., events that change the state
of the order book at the best bid or best ask levels.
More precisely we consider the following 8 dimensional counting process:
$$N_t = (\PA_t,\PB_t,\TA_t,\TB_t,\LA_t,\LB_t,\CA_t,\CB_t)$$
where:
\begin{itemize}
\item $\PA$ (resp. $\PB$) counts the number of upward (resp. downward) mid-price moves.
\item $\TA$ (resp. $\TB$) counts the number of market orders at the ask (resp. bid) that do not move the price.
\item $\LA$ (resp. $\LB$) counts the number of limit orders at the ask (resp. bid) that do not move the price.
\item $\CA$ (resp. $\CB$) counts the number of cancel orders at the ask (resp. bid) that do not move the price.
\end{itemize}

\begin{remark}
Let us stress that mid-price moves can correspond to the occurrence of a market order or a cancel order that eats all the available liquidity at best bid or best ask or to a limit order placed in the spread between best bid and best ask. We choose to not distinguish these events in order to handle a process with relatively low dimension.
\end{remark}

\noindent The previous counting process can be considered as an 8-dimensional Hawkes process characterized by $8$ exogenous intensities and $64$ kernels. We denote by $\phi^{N^j\rightarrow N^i}$ the kernel $\phi^{ij}$ coding the influence of the $j^{th}$ process on the $i^{th}$ intensity. For example $\phi^{\TB\rightarrow \PA}$ corresponds to the influence of the trades at the bid that left the mid-price unchanged on the upward price moves. These intensities and kernels can be estimated from empirical book data using the procedure described in Section \ref{thealgo}. \\

\noindent Notice that, in the following, we never impose ``by hand'' any (bid,downward)-(ask,upward) symmetry. However, it can be seen in Appendix \ref{plot} that, as expected, this symmetry is rather well satisfied (for example, $\phi^{\TB\rightarrow \PA}\simeq \phi^{\TA\rightarrow \PB}$) in our estimations. This indicates that, as suggested on simulations, while we perform our kernel estimations on processes with many dimensions and on many time scales, our method and the amount of data that we use, allows us to retrieve significant results about the kernels. Indeed, an unstable method would hardly allows one 
to retrieve this symmetry.

\subsection{Description of the database}
\label{data}
The financial data used in this paper have been provided by the company QuantHouse EUROPE/ASIA (http://www.quanthouse.com). It consists in all level-I order book data\footnote{That is the times of market orders and limit and cancel orders at the first limit and the state of the first bid-ask queues at these times.} of BUND and DAX future contracts. For every day, we only keep the most liquid maturity and we use data over one year from June 2013 to June 2014. Each file lists all the changes in the first limit (best ask or best bid) of the order book at a micro second precision. We can thus easily precisely compute from this data the times of the different market events that will be of interest here (ask or bid market, limit or cancel orders at the first limit and upward and downward mid price moves).\\

\noindent At the time scales that we shall study (from $10^{-5}$ to $10^2$ seconds), the order book dynamics strongly depend on the tick size of the asset (or more precisely by the quotient between the average spread and the tick size, see \cite{bouchaud2009markets} or \cite{dayri2012large} for characterization and differences between small and large tick assets). In that respect, the DAX corresponds to a ``small-tick'' asset while
the Bund is a ``large tick'' asset.\\

\noindent The number of events in our sample is summarized in Table \ref{ttN}.
\begin{table}[H]
\begin{center}
\begin{tabular}{|c|c|c|c|c|c|c|c|c|}
\hline
&$\PA$&$\PB$&$\TA$&$\TB$&$\LA$&$\LB$&$\CA$&$\CB$\\
\hline
xFDAX &9.36 & 9.35 & 2.66 & 2.67 & 19.7 & 19.6 & 23.3 & 23.1\\
\hline
xFGBL &1.72 & 1.72 & 3.40 & 3.48 & 29.8 & 29.8 & 26.6 & 26.5 \\
\hline
\end{tabular}
\end{center}
\caption{Total number of events (in millions).}
\label{ttN}
\end{table}

\noindent We have checked using numerical simulations of multidimensional Hawkes processes with power-law kernels, that the size of the sample is sufficient to provide reliable estimations of the shapes of the kernels. We have also checked that our main results do not depend on intraday seasonal effects: if one selects a 1 hour intraday time slice, we obtain the same results (up to some statistical noise).


\subsection{Conditional law estimations}
\label{sec:claw}
To estimate the 64 conditional laws we proceed as explained in
Section \ref{linearinterp} (see also
Appendix \ref{app_est}) taking the parameters $h_\delta=0.05$, $h_{min}=0.1$ milliseconds and $h_{max}\simeq 1059$ seconds (see \eqref{hhh})  on the period from 1 June 2013 to 1 June 2014 (252 open days).\\

\begin{figure}[H]
\centering
\includegraphics[width=\textwidth]{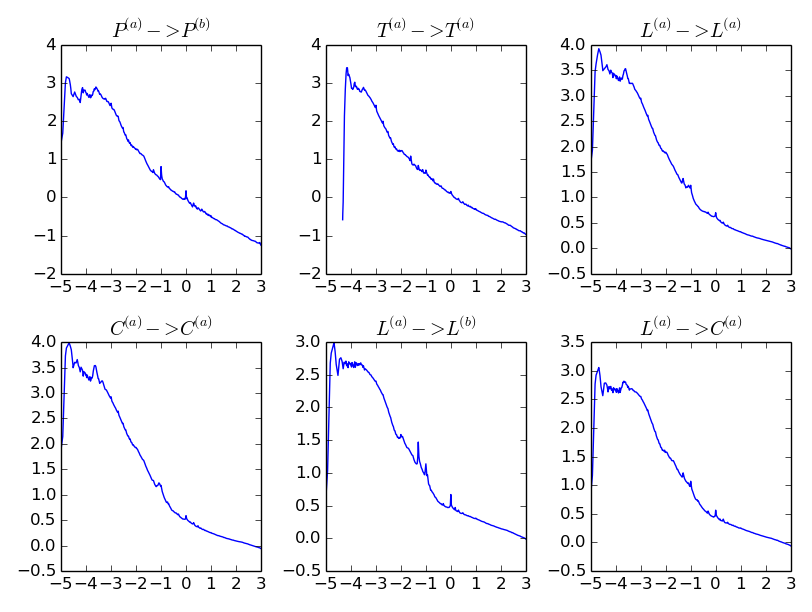}
\caption{$\log_{10}-\log_{10}$ empirical conditional laws $g^{\PA\rightarrow \PB}$, $g^{\TA\rightarrow \TA}$, $g^{\LA\rightarrow \LA}$ and $g^{\CA\rightarrow \CA}$ for the Bund.
}
\label{llclaw3}
\end{figure}

\noindent Surprisingly most of the conditional laws between financial events have very similar properties. As an illustration, Figure \ref{llclaw3} displays, for the Bund, $g^{\PA\rightarrow \PB}$, $g^{\TA\rightarrow \TA}$, $g^{\LA\rightarrow \LA}$, $g^{\CA\rightarrow \CA}$, $g^{\LA\rightarrow \LB}$ and $g^{\LA\rightarrow \CA}$ in $\log_{10}-\log_{10}$.

\noindent The first thing that we notice is that above a time scale $\tau_1\sim 0.1$ seconds, these conditional laws behave as $t^{-\gamma}$ with $\gamma<1$. For $g^{\TA\rightarrow \TA}$, this
is the well known long-range memory of the order flows. Between $\tau_1$ and $\tau_2\sim 0.3 ms$, the conditional laws roughly behaves as a power law of exponent of order $1$. Below, $\tau_2$, the conditional laws ``saturate".\\

\noindent We also observe a few ``bumps" on the conditional law which do not correspond to noise in the estimation. We believe that there are two kinds of bumps. The first kind of bumps appear at ``round" times (0.01, 0.1, 0.5, 1 and 2 seconds), we believe that they are due to the automatization of trading. For example, if an algorithm posts a limit order every second, this implies a bump in the conditional law at 1 second. The second kind of bumps appears around $\tau_2\sim 0.3$ milliseconds. We believe that it corresponds to the average reaction time (i.e., the average ``latency'') of the agents  to an event. The decreasing of the conditional law below 0.03 milliseconds is an artefact of our data.\\

\noindent Once the conditional laws have been estimated, we solve the Wiener-Hopf system, following
steps 2. and 3. in Section \ref{thealgo} with the parameters $K=100$ (so that $\delta\simeq 0.15$), $T_{min}=0.1$ milliseconds and $T_{max}\simeq 140$ seconds and compute the exogenous intensity estimates following step 4.
We shall first comment on the values of the exogenous intensities.

\subsection{Exogenous intensities}
Let us recall that, within the Hawkes model, the exogenous intensity $\mu^i$
can be interpreted as the rate of (Poisson) events of type $i$ that are ``coming'' from an exogenous source of information, i.e., that are not ``caused'' by any other past event in the model. In that respect, the ratio:
\begin{equation}
\label{ratio}
R^i = \frac{\mu^i}{\Lambda^i}
\end{equation}
represents an exogeneity ratio, namely the ratio between the number of exogenous
events and the total number of events of type $i$.
Notice that, in the one dimensional case, according to Proposition \ref{order1}, one has simply:
$R = 1-||\phi||_1$.

\begin{table}[H]
\begin{center}
\begin{tabular}{|c|c|c|c|c|c|c|c|c|}
\hline
&$\PA$&$\PB$&$\TA$&$\TB$&$\LA$&$\LB$&$\CA$&$\CB$\\
\hline
$\mu$ &2.37e-2 & 2.39e-2 & 1.06e-2 & 1.14e-2 & 2.07e-2 & 2.27e-2 & 1.53e-2 & 7.51e-3 \\
\hline
$R $&2.7\% & 2.7\% & 4.3\% & 4.5\% & 1.1\% & 1.2\% & 0.7\% & 0.4\% \\
\hline
\end{tabular}
\end{center}
\caption{Estimated exogenous intensities (in $s^{-1}$) and the corresponding exogenous ratio \eqref{ratio} of the DAX futures.}
\label{tt1}
\end{table}

\begin{table}[H]
\begin{center}
\begin{tabular}{|c|c|c|c|c|c|c|c|c|}
\hline
&$\PA$&$\PB$&$\TA$&$\TB$&$\LA$&$\LB$&$\CA$&$\CB$\\
\hline
$\mu$&7.13e-3 & 7.10e-3 & 1.41e-2 & 1.45e-2 & 3.83e-2 & 3.83e-2 & 4.00e-2 & 4.39e-2 \\
\hline
$R $&4.4\% & 4.4\% & 4.5\% & 4.5\% & 1.4\% & 1.4\% & 1.6\% & 1.8\% \\
\hline
\end{tabular}
\end{center}
\caption{Estimated exogenous intensities (in $s^{-1}$) and the corresponding exogenous ratio \eqref{ratio} of the BUND futures.}
\label{tt2}
\end{table}
\noindent The estimated intensities and the corresponding exogenous ratio for the DAX and the BUND futures are reported
respectively in Tables \ref{tt1} and \ref{tt2}.
These tables reveal that, for both assets, the level of exogeneity is very low:
$R$ is only a few percent, meaning that most of
the events can be considered to be directly triggered by past events within this model.
See \cite{jaisson2013limit,jaisson2014fractional} for theoretical studies of the diffusive long term behavior of such almost purely endogenous Hawkes processes.
For comparison, a simple one dimensional Hawkes model accounting for mid-price jump events as in refs \cite{filimonov2012quantifying} provides an exogenous ratio of respectively $R = 0.059$ for the DAX and $R = 0.10$ for the BUND future. This means that the 8-dimensional Hawkes model provides a better description of mid-price changes than a simple one dimensional model that has to involve a larger amount of ``external'' sources of information.\\

\noindent Let us also remark that, for both assets, market orders are more exogenous than limit and cancel orders. This is not surprising since several studies tend to show that market orders are ``leading'' limit and cancel.  Thus, degree of endogeneity of the limit and cancel orders should be greater than the one of the market orders.\\

\noindent Finally, one can see that though limit and market orders have strikingly
similar exogenous ratio for small and large tick assets, the mid-price changes
and cancel occurrences are more endogenous in the case of small tick asset (DAX).

\subsection{Matrix of kernel norms $||\phi||_1$}
\begin{figure}[H]
\centering
\includegraphics[width=\textwidth]{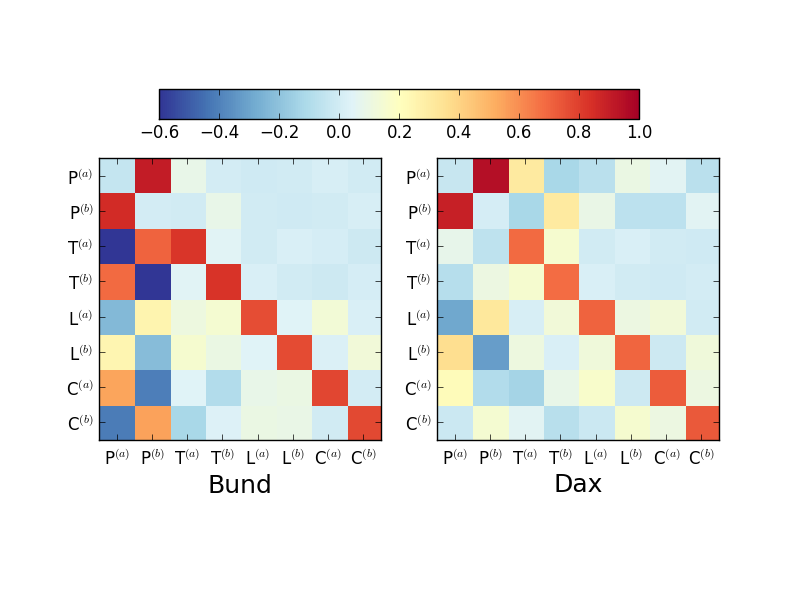}
\vspace{-59pt}
\caption{The matrix of estimated $||\phi^{ij}||$.}
\label{normphi}
\end{figure}

Before discussing the precise shape of the estimated kernels, we use the values of the norms of the kernels $\phi^{ij}$ to comment the main mutual and self excitations that occur in the order book\footnote{Let us recall that the norm $||\phi^{ij}||_1$ represents the mean number of events of type $i$ triggered by an event of type $j$.}.
The matrices of the kernel norms $||\phi||_1$ we obtained for the DAX and the BUND are reported in Figure \ref{normphi}.
For the sake of simplicity, we have represented the norm values using a colormap from blue to red. Notice that blue values correspond to negative kernel norms\footnote{Let us remark that we allow an abuse of language in the sense that $||\phi||_1$ stands for $\int_0^\infty \phi(s) ds$ which is not really a norm unless $\phi(t) > 0$.}.
This feature has a natural interpretation as an inhibitory
effect within the non-linear
version of the Hawkes model described by Equation \eqref{defHawkes2}.
We have checked, on numerical simulations, that as long as the realized quantity \eqref{defHawkes1} remains most of the time positive, the non parametric estimation procedure introduced in Section \ref{thealgo}, when applied to the non linear Hawkes process \eqref{defHawkes2}, leads to reliable estimations even for negative kernels.\\

\noindent As expected, the overall symmetry (ask/bid) is fairly well recovered empirically on the matrix shape.
Any $2\times 2$ sub-matrix with homogeneous inputs (i.e., same type of inputs to be chosen among Price changes, Trades, Cancels or Limits) and homogeneous outputs is symmetric.
Thus for instance the kernel $\phi^{\TA \rightarrow \CB}$ seems to have the same norm as the kernel
$\phi^{\TB \rightarrow \CA}$. Let us point out that, since the Dax is a stock index we could have expected some discrepancy between the ask side and the bid side (statistics of downward jumps are slightly different than statistics of upward jumps on equity markets) however, this discrepancy is negligible on intraday data.\\

\noindent One striking feature that clearly appears on both matrices is the anti-diagonal shape of the $(\PA,\PB)$ sub-matrix and the diagonal shape of the $(\TA,\TB)$, $(\LA,\LB)$, $(\CA,\CB)$ sub-matrices, see also Paragraphs \ref{PP}, \ref{TT}, \ref{LL} and \ref{CC} in appendix where the corresponding normalized cumulated kernels are plotted. This means that, on the one hand, market, limit and cancel orders mainly cause events of the same type and sign. This property can be mainly attributed to the splitting of metaorders, see \cite{lillo2005theory}, and to a less extent to some herding behaviour of agents. We will see in the next section that these diagonal kernels are well described by roughly power-law decreasing functions. As far as mid-price jumps are concerned, the cross-exciting structure between $\PA$ and $\PB$ implies a strongly mean-reverting behaviour, which is of the main characteristic of the price microstructure and  which somehow guarantees the absence of long range correlation in prices, see \cite{bacry2014hawkes}, in agreement with market efficiency. Note that this cross price kernel can be linked to the kernel of the propagator model, see \cite{bouchaud2004fluctuations,jaisson2014market}. \\

\noindent Finally, let us point out that that price changes appear to be the components that influences the most the other components, i.e., a price change ``drives'' the dynamics more than anything else.
We will come back on this feature and on other features of this matrix (impact of orders on the price, impact of the price on events, impact of trades on the liquidity,...) in the following sections.

\subsection{Matrix of norms $||\psi||_1$}
\begin{figure}[H]
\centering
\includegraphics[width=\textwidth]{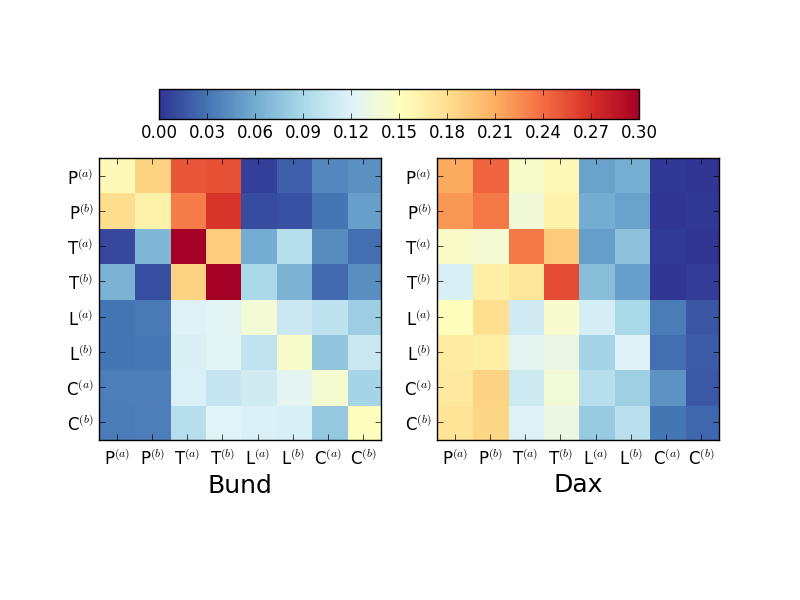}
\vspace{-59pt}
\caption{The matrix of estimated $||\bar{\psi}^{ij}||$.}
\label{normpsitilde}
\end{figure}

Let us recall that the kernel $\phi^{ij}$ can be considered
as the ``bare'' impact of an event of type $j$ on an event of type $i$. If one wants to account for the ``dressed'' impact, i.e. the impact associated with all the cascade
triggered by some event, one has to estimate the function $\psi^{ij}$ as defined by \eqref{defpsi} whose matrix of norms can be computed as:
$$||\psi||_1=\||\phi||_1(\mathbb{I}-||\phi||_1)^{-1}.$$
More precisely, if one introduces:
$$||\bar{\psi}^{ij}||_1 = \frac{\mu^j}{\Lambda^i} ||\psi^{ij}||_1$$
then, according to the population interpretation mentionned in section \ref{sec:secondorder} and Proposition \ref{order1}, $||\bar{\psi}^{ij}||$ corresponds to the fraction of events of type $i$ triggered by {\em exogenous} events of type $j$ (while $||\psi||_1$ corresponds to the average number of events of type $i$ indirectly generated by an event of type $j$).
The matrix norm $||\bar{\psi}||$ is displayed in Figure \ref{normpsitilde}.

\noindent One can see that exogenous limit and cancel orders have a poor
influence on mid-price variations that are mainly caused
by exogenous trades and price jumps. This indicates that exogenous information is mainly incorporated into
prices through trades or orders that directly shift the mid-price. One
can also remark that in the case of large tick asset (BUND), an exogenous price move mainly impacts the price itself but not the order flow that does not move
the price while for the small tick asset (DAX), an exogenous
price variation impacts all type of events.


\subsection{Shape of the kernels}
\label{shape}
In all former studies where non parametric estimations of Hawkes kernels involved in
the dynamics of order flows
were performed \cite{bacry2012non,bacry2014hawkes,hardiman2013critical}, power-law kernels with exponents close to 1 have been observed.
This property can be directly linked to the strong persistence of the order flow dynamics, see \cite{jaisson2014market}, mainly caused by the splitting of large orders and, to a lesser extent,  to the herding behavior of agents. This feature remains true in the 8-dimensional description adopted in this paper when one focuses on the kernels involved in the self-excitation of market, limit and cancel orders and in the cross excitation of mid-price moves (the kernels with highest norms, see Figure \ref{normphi}).\\

\begin{figure}[H]
\centering
\includegraphics[width=\textwidth]{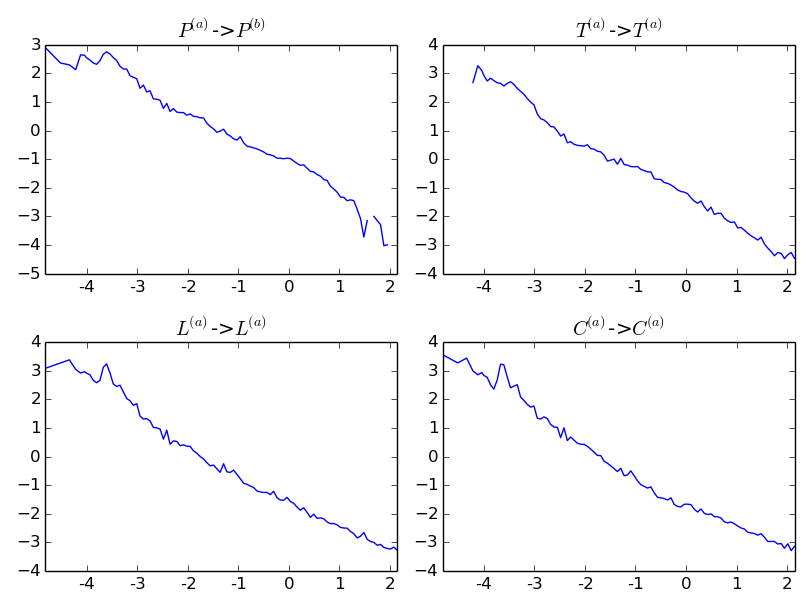}
\vspace{-32pt}
\caption{$\log_{10}-\log_{10}$ plots of $\phi^{\PA\rightarrow \PB}$, $\phi^{\TA\rightarrow \TA}$, $\phi^{\LA\rightarrow \LA}$ and $\phi^{\CA\rightarrow \CA}$. These kernels display loose power-law behavior with an exponent close to 1. For illustration purposes, fits were performed on the entire time-scale range, though they clearly display different regime.}
\label{loglogpowerlaw}
\end{figure}

\noindent Figure \ref{loglogpowerlaw} displays $log_{10}-log_{10}$ plots of the kernels $\phi^{\TA\rightarrow \TA}$, $\phi^{\LA\rightarrow \LA}$, $\phi^{\CA\rightarrow \CA}$ and $\phi^{\PA\rightarrow \PB}$. It clearly appears that these kernels loosely behave as a power law of exponent slightly higher than one.

\noindent Let us stress that this ``roughly power law behavior'' is not true for all kernels. For example, the kernels $\phi^{\TA\rightarrow \PA}$ and $\phi^{\PA\rightarrow \LA}$ are ``localised'' below the millisecond, see Figure \ref{loc} and the cumulated kernels in Appendix \ref{plot}. Empirical kernels have a much ``richer'' behavior than conditional laws which as stated earlier all have the same stylized facts.
\begin{figure}[H]
\centering
\includegraphics[width=\textwidth,height=55mm]{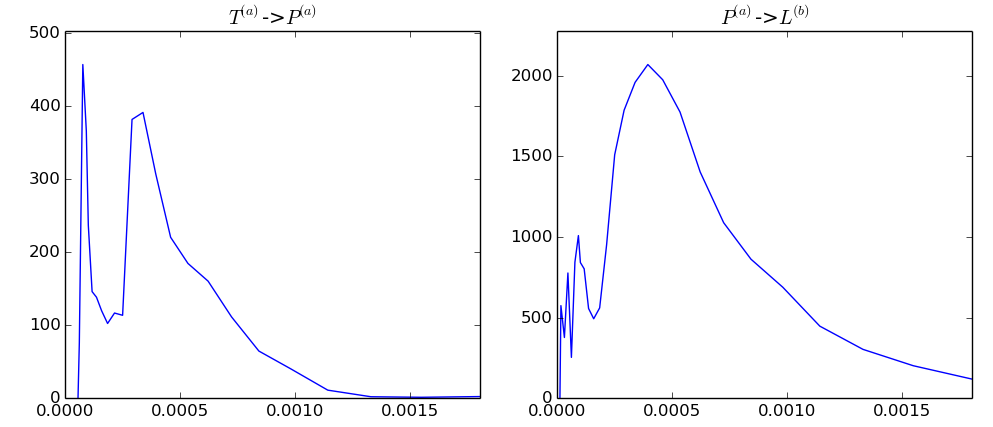}
\vspace{-18pt}
\caption{$\phi^{\TA\rightarrow \PA}$ and $\phi^{\PA\rightarrow \LA}$ as a function of time below the millisecond.}
\label{loc}
\end{figure}

\subsection{The impact of the order flows on the price}

For the sake of clarity, we split the two $8 \times 8$ matrices of kernels into $2\times 2$ sub-matrices.  This representation of the kernels is provided in Appendix \ref{plot} for both DAX and BUND.
Each sub-plot represents the normalized cumulated kernels
$\frac{\Lambda^j}{\Lambda^i}\int_0^t \phi^{ij}(u) du$ as a function of $\log_{10}(t)$  (so there are $2\times 64$ of them).
This normalization for the kernel is natural in the population dynamics interpretation of Hawkes processes, see \cite{hawkes1974cluster}. Indeed, $\frac{\Lambda^j}{\Lambda^i}||\phi^{ij}||$ corresponds to the proportion of $i$ whose parent is a $j$ (while $||\phi^{ij}||$ corresponds to the average number of children of type $i$ for an individual of type $j$).\\

\noindent In this section, we focus on the impact of the flow of orders on the price, i.e., on the events which do not instantaneously move the price and that have an impact on mid-price variations.
It corresponds to Paragraphs \ref{TP}, \ref{LP} and \ref{CP} in appendix.\\

\noindent One can see that the trades have a very localized delayed {\em self} impact on price variations (by {\em self} we mean {\em in the same direction}, e.g., an ask market order
implies an eventual upward price move). Indeed, it seems that it takes 0.3 milliseconds for the market to incorporate the information of the trade. This is the averaged time for other agents to react to this order (this corresponds to the ``latency'' delay we have observed on the conditional laws in Section \ref{sec:claw} and the kernels in Section \ref{shape}).
After this short time scale, the kernel is negligible, that is the cumulated kernel remains constant, and the trade does not imply any more price changes.\\

\noindent As far as limit and cancel orders are concerned, their impacts appear to be less localized and to be different when one compares the DAX (small tick)
and the BUND (large tick). For the DAX, limit orders impact the price by exiting the price moves in the opposite direction (limit orders at the bid trigger upward price moves and conversely) and by slightly inhibiting
the price moves in the same direction. For the BUND, limit orders mainly have an inhibitory effect one price changes (limit orders at the bid inhibit downward price moves). As opposed to the impact of market orders, the impact of limit orders is not very fast. While the information of market orders is almost immediate (0.3 milliseconds), a limit order needs time to be significant. \\

\noindent Cancel orders impact the price in a very similar way as limit orders but in the opposite direction.


\subsection{The impact of price jumps on the order flows}
The kernels that correspond to the impact of price jumps on the order flows correspond to Paragraphs \ref{PT}, \ref{PL} and \ref{PC} where the normalized cumulated kernels\footnote{See beginning of the previous section or of Appendix \ref{plot}} are plotted.\\

\noindent The most striking feature is that the effect of a mid-price change on market orders strongly depends on the tick size (as it is also observed on Figure \ref{normphi}). For the DAX (small tick), this effect is rather small and {\em self-exciting} (upward price moves imply ask trades). On the contrary, for the BUND (large tick), this effect is far more important and is {\em cross-exciting} (upward price moves trigger bid trades and inhibit ask trades). This can be interpreted in terms of adverse selection: On the one hand, for large tick assets, after an upward price move, agents will not want to execute market orders at the new ask price because this price will be too high. On the other hand , the new bid price will be more interesting and agent will thus execute bid market orders.\\


\noindent On both assets, the influence of price moves on the liquidity is mostly {\em self} in the sense that upward price moves excite bid limit orders and ask cancel orders (and inhibit ask limit orders and bid cancel orders). The excitations are mostly localized (around 0.3 milliseconds) and correspond to the averaged time reactions of market makers to the new information while the inhibition kernels are long term. One can also notice that for the DAX, the short term exciting effect of say $\phi^{\PA\rightarrow \CA}$ is balanced by a longer term inhibitory effect (around 10 seconds). We believe that this effect is linked to the reversion of the liquidity to its stationary state a ``long'' time after the price change has occurred.

\subsection{The impact of market orders on liquidity and vice-versa}
The kernels that correspond to the impact of the market order flow on limit and cancel order flows are in Paragraphs \ref{TL} and \ref{TC} where the normalized cumulated kernels are plotted.\\

\noindent For both assets, trades have very fast positive influence on the opposite limit orders. This is due to the fact that the underlying ``efficient'' price has moved and the liquidity must adapt itself. At longer time scales, the effect is reversed (the cross kernels become negative, i.e., trades inhibit cross limit orders). This is due to the reversion of the liquidity towards its stationary state. As far as the influence of market orders on the rate of cancel orders is concerned, for the BUND, trades imply cancel orders in the same direction at very short time scale and then, at longer time scales, this influence becomes negative. This effect is very small which means that the proportion of limit and cancel orders directly implied by a trade is small. However, we see on Figure \ref{normpsitilde} that when we ``dress'' this impact with intermediate events, exogenous trades indirectly imply a significant proportion of limit and cancel orders.\\

\noindent Concerning the reverse impact of limit and cancel order flows on market order flow, the corresponding kernels are displayed in Paragraphs \ref{LT} and \ref{CT}. One can see that ask (resp. bid) cancel and bid (resp. ask) limit orders excite the trading rate at the ask (resp. the bid).

\subsection{The impact of the limit order flow on the cancel order flow and vice-versa}
The kernels that correspond to the dynamics between limit and cancel order flows correspond to Paragraphs \ref{LL}, \ref{CC}, \ref{LC} and \ref{CL} which display the corresponding normalized cumulated kernels.\\

\noindent As for market orders, the limit/limit kernels and the cancel/cancel kernels are mostly self-exciting (e.g., $\phi^{\LA\rightarrow \LA}$ is dominant and $\phi^{\LA\rightarrow \LB}$ is negligible). Remark also that cancel orders have a short term influence on cross limit orders ($\phi^{\CA\rightarrow \LB}$ is localized around 0.3 milliseconds) which corresponds to the information of the cancel order and a longer term impact on self limit orders ($\phi^{\CA\rightarrow \LA}$ is a power law) which corresponds to the return of the liquidity. \\

\noindent In a similar way, limit orders have a short term influence on cross cancel orders and a longer term impact on self cancel orders.

\section{Discussion and concluding remarks}
\label{conc}


We have presented a numerical solution of the Wiener-Hopf Equation \eqref{wh} when some kernels behave as power laws of exponents slightly higher than one and are thus significant over a very wide range of time scales (from 100 microseconds to 100 seconds!) and in a rather large number of dimensions. This lead us to an efficient algorithm for non-parametric estimation of these kernels.\\

\noindent Using the natural causal interpretation of Hawkes processes, we applied this algorithm to the study of
high frequency financial data (timestamped with  a precision of a micro-second). It allowed us to disentangle the cross-influences between eight types of first limit events occurring in order books.\\

\noindent Our approach allowed us to retrieve some well known stylized facts about market dynamics:
\begin{itemize}

\item{The order flows are strongly self excited: The main influence of market, limit and cancel orders is on themselves. This is linked to the well known persistence of order flows and to the splitting of meta-orders into sequences of orders. Moreover, orders also have an influence of orders of the same ``direction''. For example, ask trades excite bid limit orders and ask cancel orders. The same kind of behavior had already been observed in \cite{eisler2012price} by directly looking at the correlation functions between these events.}

\item{The prices are efficient: To balance the strong persistence of the order flow imbalance, the influence of price changes on themselves is mostly cross. That is upward price moves cause downward price moves. This is consistent the measure of the ``bare propagators'' of \cite{eisler2012price}. Indeed, in \cite{eisler2012price}, the bare impact of events on the price is shown to decrease in time. Not however that this bare impact is not exactly equivalent to our kernels since it does not fully ``disentangle'' the influences of all the events on all the events but only the influence of orders on price changes.}

\item{The orders impact the price even if they do not move it mechanically. Again, this is consistent with the measures of the propagators, of \cite{eisler2012price}, of the events that do not move the price.}

\end{itemize}

\noindent The generality of our Hawkes model allowed us to account for much richer dynamics than previous works and to describe and quantify the influences between all types of events. We have thus found some new and more subtle results that to our knowledge had never been observed.
\begin{itemize}

\item{Price moves have a retro influence on orders flows: These are the kernels that the tick size influences the most. For large tick assets, adverse selection prevails and this effect is cross: upward price moves excite bid trades and vice versa. For small tick assets, this effect is self because of the persistence of order flows.}

\item{The Hawkes framework and our estimation of the kernels allows us not only to have a measure of ``causality'' between the different events but also of the time scales at which this causality appears. We have found that there are loosely two kinds of influences: Fast and localized influences, for which the volume of the kernel is localized around the reaction time of the market (on our data 0.3 milliseconds) and influences which are spread over a wide range of time scales and whose corresponding kernels thus behave somehow as power law functions of exponents slightly higher than one. }

\item{Note also that as opposed to \cite{eisler2012price}, our model is a real time model and not a discrete time model where each period corresponds to an event. This allows us for example, to see that market and limit orders have different impacts. Market orders have localized impacts while limit orders take time to have an impact.}

\item{Finally, we can measure the endogeneity of all market events. We find in particular that market orders are more exogenous than limit and cancel orders. Therefore, when looking at the ``dressed kernels'' $\psi$, we find that the trades are leading in the sense that, exogenous trades have more influence than exogenous limit and cancel orders.}

\end{itemize}


\noindent One thing that probably lacks in our purely Hawkes order book model is that the arrival of events does not directly depend on the state of the order book but only on the past events. For example, it is not clear that in our model, when the spread is important, liquidity will arrive. On the other side of the spectrum of order book models, we can mention \cite{gareche2013fokker,huang2013simulating} where the order book dynamics are purely Markovian and thus do not depend on the history of the order flow. We believe that ``reality'' is between these two approaches: the intensity of market events depends both on their history and on the state of the order book.\\


\noindent The stability and tractability of our numerical estimation method over extremely wide ranges of time scales and as the dimension of the model increases imply that we have many leads to improve our order book model. For example, we can add marks to account for the size of the different events, see \cite{bacry2014second} or exogenous terms to account for external news, see \cite{rambaldi2014modeling}.\\

\noindent We conclude by underlying that of course, the relevance of Hawkes models is not restricted to finance. Therefore, our procedure can be applied to the study of any field where the measure of complex relations between point processes is necessary.



\section*{Acknowledgements}

\noindent 
We thank M.~Rosenbaum for useful discussions.
This research benefited from the support of the Chair Markets in Transition, under the aegis of Louis Bachelier Finance and Sustainable Growth laboratory, a joint initiative of Ecole
Polytechnique, Universit\'e d'Evry Val d'Essonne and F\'ed\'eration Bancaire Fran\c{c}aise.

\appendix
\section{Multiscale estimation of the conditional laws}
\label{app_est}
The aim of this appendix is to present our multi scale estimation procedure of the conditional law $g$ defined by \eqref{def_claw}.

\subsection{Estimation procedure of $g$}

Let us assume that, for all $(i,j)$, we have the times of the events $i$ $(T_k^i)_{k\leq n_i}$ and $j$ $(T_k^j)_{k\leq n_j}$ and we want an empirical estimation of the conditional laws $g^{ij}$.\\

\noindent
To do that, we will choose a time grid $(t_l)_{l\leq n}$ small enough so that we can approximate $g^{ij}(\frac{t_l+t_{l+1}}{2})$ by $\frac{1}{t_{l+1}-t_l}\int_{t_l}^{t_{l+1}} g^{ij}(s)ds$ but large enough so that there is enough points $(k,k')$ such that $T_k^i-T_{k'}^j\in [t_{l},t_{l+1}]$ to have a good approximation of $\int_{t_l}^{t_{l+1}} g^{ij}(s)ds$ by $\frac{1}{n_j} \sum_{k=1}^{n_i} \sum_{k'=1}^{n_j}\mathbbm{1}_{T_k^i-T_{k'}^j\in [t_{l},t_{l+1}]}$. We thus approximate $g^{ij}(\frac{t_l+t_{l+1}}{2})$ by $$\hat{g}^{ij}(\frac{t_l+t_{l+1}}{2})=\frac{1}{t_{l+1}-t_l}\frac{1}{n_j} \sum_{k=1}^{n_i} \sum_{k'=1}^{n_j}\mathbbm{1}_{T_k^i-T_{k'}^j\in [t_{l},t_{l+1}]}$$
and we take a piecewise affine interpolation of $\hat{g}^{i,j}$ between the points $\frac{t_l+t_{l+1}}{2}$.

\subsection{Choice of the grid}

The first natural grid that one can take is a uniform grid: $[0,h,2h,...,h_{max}]$. However, if one does that, one cannot get a good estimation of the conditional law on a large range of time scales. For example, if one takes $h=0.0001s$ to have a good estimation of the conditional law at low time scales, one cannot estimate get a good estimation around one second. Indeed, there will be very few points of $i$ between 1 and 1.0001 seconds after a point of $j$.\\

\noindent To solve this problem, we will consider a grid that is uniform between 0 and $h_{min}$ and log-uniform between $h_{min}$ and $h_{max}$:
\begin{equation}
\label{hhh}
[0,h_{min}h_\delta,h_{min}2h_\delta,... ,h_{min},h_{min} e^{h_\delta},h_{min} e^{2h_\delta},...,h_{max}].
\end{equation}
Doing this, we will have enough points between two points of the grid but the claw will not vary too much between two points of the grid.

\section{Estimation results for the full order book model: the cumulated kernel matrix $\int_0^t \phi(s) ds$}
\label{plot}
In this Appendix, we represented the resulting kernels estimated on the high frequency database (see Section \ref{s3}).
For each $i$ and $j$ in $$\{ \PA, \PB, \TA, \TB, \LA, \LB, \CA, \CB \}$$ we have represented
the estimated normalized cumulated kernels
$\frac{\Lambda^j}{\Lambda^i}\int_0^t \phi^{i \rightarrow j}(u) du$ as a function of $\log_{10}(t)$.
This normalization for the kernel is natural because in the population dynamics interpretation of Hawkes processes $\frac{\Lambda^j}{\Lambda^i}||\phi^{i\rightarrow j}||_1$ corresponds to the proportion of $i$ whose parent is a $j$ (while $||\phi^{i\rightarrow j}||_1$ corresponds to the average number of children of type $i$ for an individual of type $j$).\\

\noindent There are $2\times 64$ plots that are organized by assets and by types of events.


\captionsetup[subfigure]{labelformat=empty}

\subsection{Influence of price moves on price moves}
\label{PP}
\begin{figure}[H]
        \begin{subfigure}[b]{0.45\textwidth}
                \includegraphics[width=\textwidth,height=40mm]{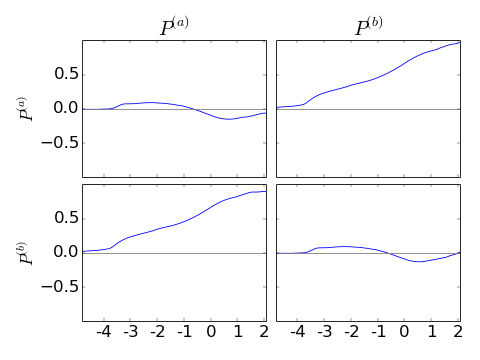}
                \caption{xFDAX}
        \end{subfigure}
        \begin{subfigure}[b]{0.45\textwidth}
                \includegraphics[width=\textwidth,height=40mm]{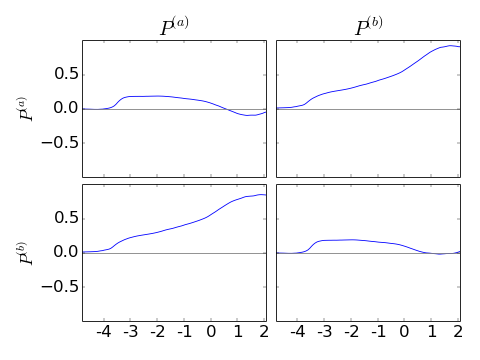}
                \caption{xFGBL}
        \end{subfigure}
\end{figure}

\subsection{Influence of trades on trades}
\label{TT}
\begin{figure}[H]
        \begin{subfigure}[b]{0.45\textwidth}
                \includegraphics[width=\textwidth,height=40mm]{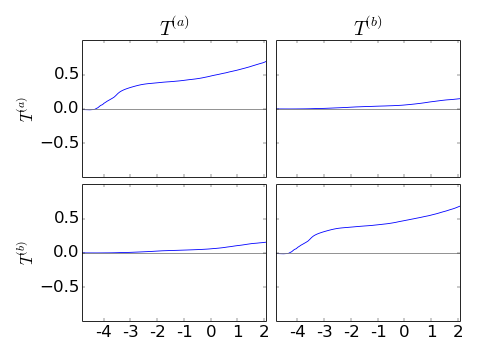}
                \caption{xFDAX}
        \end{subfigure}
        \begin{subfigure}[b]{0.45\textwidth}
                \includegraphics[width=\textwidth,height=40mm]{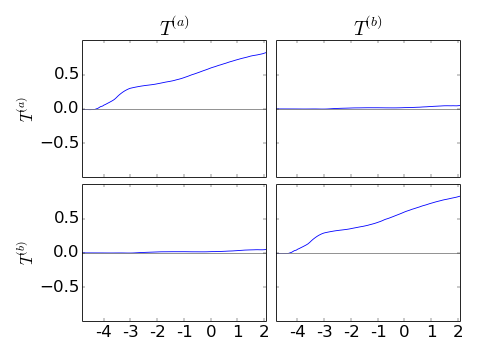}
                \caption{xFGBL}
        \end{subfigure}
\end{figure}

\subsection{Influence of limit orders on limit orders}
\label{LL}
\begin{figure}[H]
        \begin{subfigure}[b]{0.45\textwidth}
                \includegraphics[width=\textwidth,height=40mm]{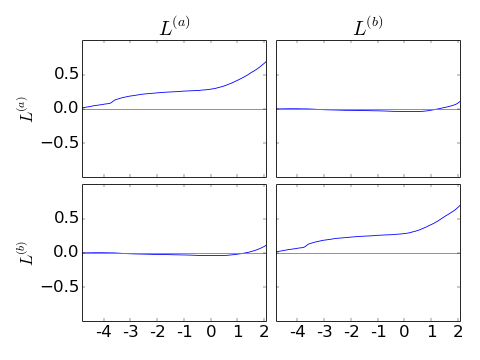}
                \caption{xFDAX}
        \end{subfigure}
        \begin{subfigure}[b]{0.45\textwidth}
                \includegraphics[width=\textwidth,height=40mm]{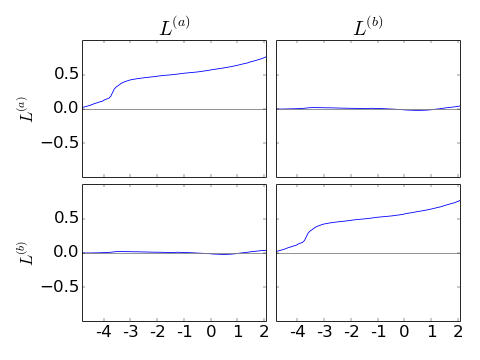}
                \caption{xFGBL}
        \end{subfigure}
\end{figure}

\subsection{Influence of cancel orders on cancel orders}
\label{CC}
\begin{figure}[H]
        \begin{subfigure}[b]{0.45\textwidth}
                \includegraphics[width=\textwidth,height=40mm]{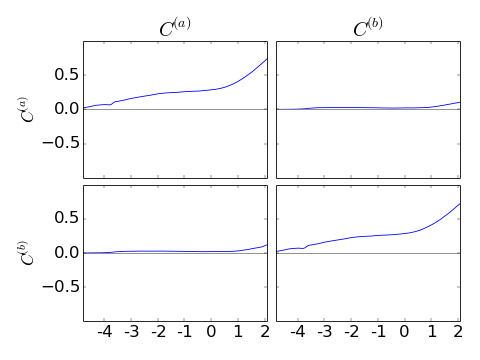}
                \caption{xFDAX}
        \end{subfigure}
        \begin{subfigure}[b]{0.45\textwidth}
                \includegraphics[width=\textwidth,height=40mm]{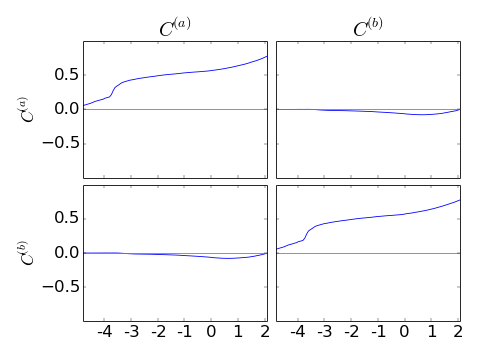}
                \caption{xFGBL}
        \end{subfigure}
\end{figure}

\subsection{Influence of trades on price changes}
\label{TP}
\begin{figure}[H]
        \begin{subfigure}[b]{0.45\textwidth}
                \includegraphics[width=\textwidth,height=40mm]{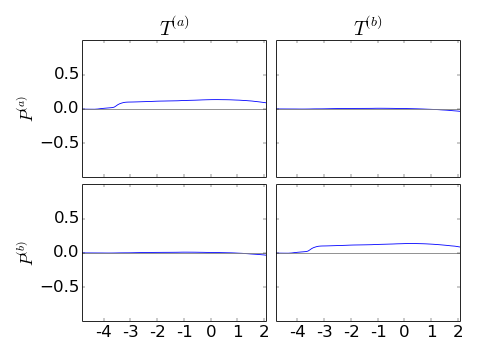}
                \caption{xFDAX}
        \end{subfigure}
        \begin{subfigure}[b]{0.45\textwidth}
                \includegraphics[width=\textwidth,height=40mm]{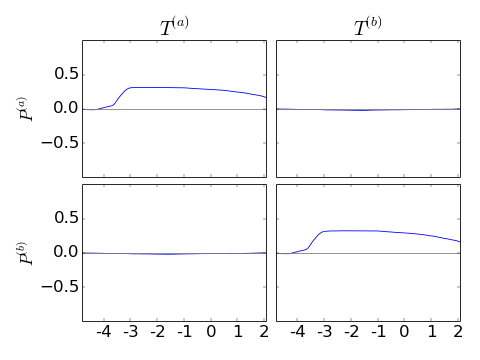}
                \caption{xFGBL}
        \end{subfigure}
\end{figure}

\subsection{Influence of limit orders on price changes}
\label{LP}
\begin{figure}[H]
        \begin{subfigure}[b]{0.45\textwidth}
                \includegraphics[width=\textwidth,height=40mm]{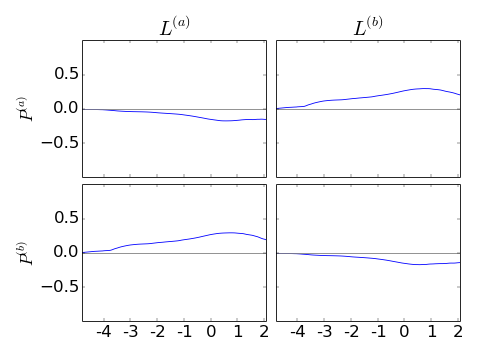}
                \caption{xFDAX}
        \end{subfigure}
        \begin{subfigure}[b]{0.45\textwidth}
                \includegraphics[width=\textwidth,height=40mm]{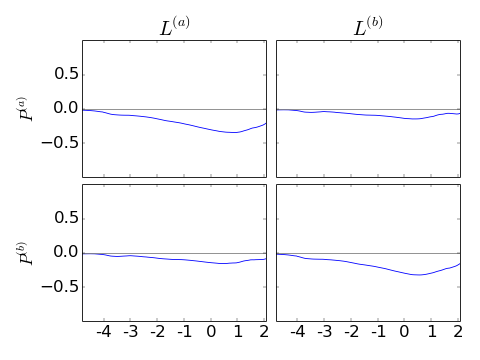}
                \caption{xFGBL}
        \end{subfigure}
\end{figure}

\subsection{Influence of cancel orders on price changes}
\label{CP}
\begin{figure}[H]
        \begin{subfigure}[b]{0.45\textwidth}
                \includegraphics[width=\textwidth,height=40mm]{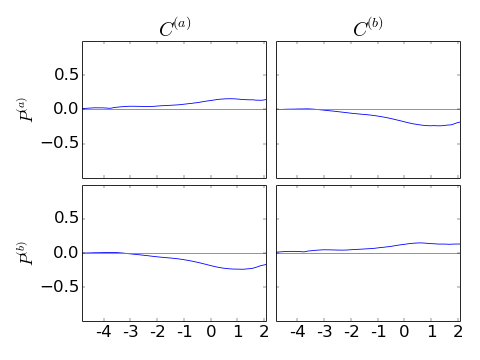}
                \caption{xFDAX}
        \end{subfigure}
        \begin{subfigure}[b]{0.45\textwidth}
                \includegraphics[width=\textwidth,height=40mm]{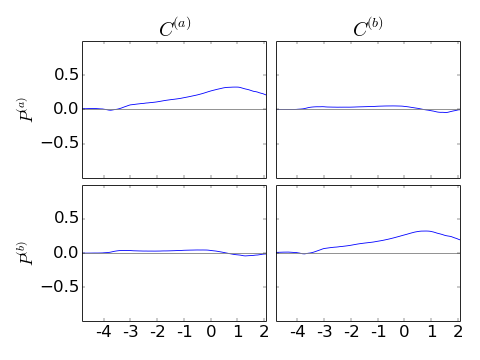}
                \caption{xFGBL}
        \end{subfigure}
\end{figure}

\subsection{Influence of price changes on trades}
\label{PT}
\begin{figure}[H]
        \begin{subfigure}[b]{0.45\textwidth}
                \includegraphics[width=\textwidth,height=40mm]{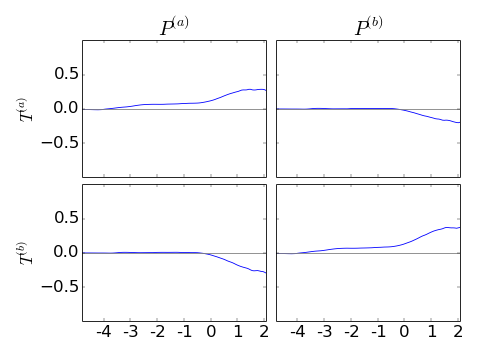}
                \caption{xFDAX}
        \end{subfigure}
        \begin{subfigure}[b]{0.45\textwidth}
                \includegraphics[width=\textwidth,height=40mm]{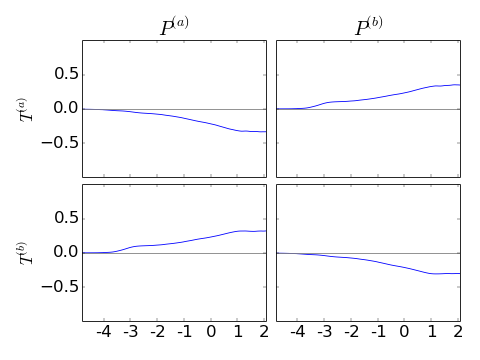}
                \caption{xFGBL}
        \end{subfigure}
\end{figure}

\subsection{Influence of price changes on limit orders}
\label{PL}
\begin{figure}[H]
        \begin{subfigure}[b]{0.45\textwidth}
                \includegraphics[width=\textwidth,height=40mm]{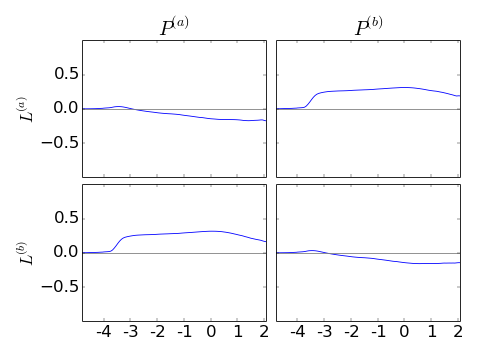}
                \caption{xFDAX}
        \end{subfigure}
        \begin{subfigure}[b]{0.45\textwidth}
                \includegraphics[width=\textwidth,height=40mm]{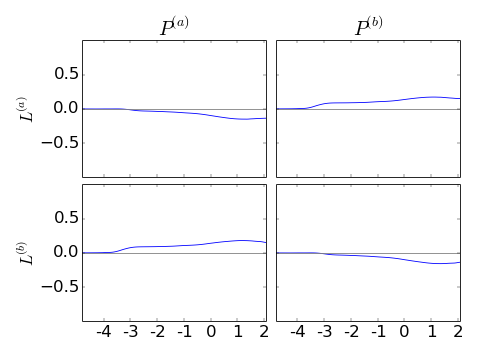}
                \caption{xFGBL}
        \end{subfigure}
\end{figure}

\subsection{Influence of price changes on cancel orders}
\label{PC}
\begin{figure}[H]
        \begin{subfigure}[b]{0.45\textwidth}
                \includegraphics[width=\textwidth,height=40mm]{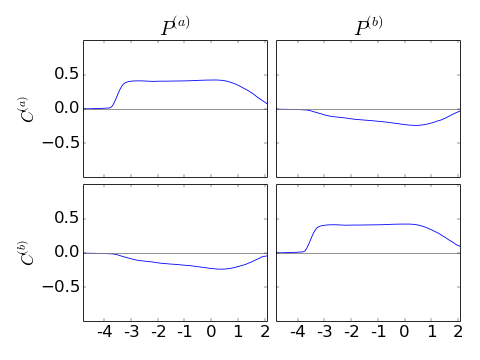}
                \caption{xFDAX}
        \end{subfigure}
        \begin{subfigure}[b]{0.45\textwidth}
                \includegraphics[width=\textwidth,height=40mm]{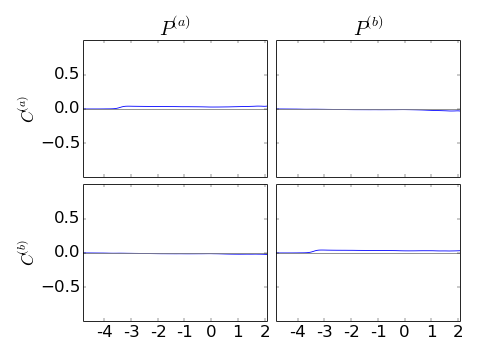}
                \caption{xFGBL}
        \end{subfigure}
\end{figure}

\subsection{Influence of trades on limit orders}
\label{TL}
\begin{figure}[H]
        \begin{subfigure}[b]{0.45\textwidth}
                \includegraphics[width=\textwidth,height=40mm]{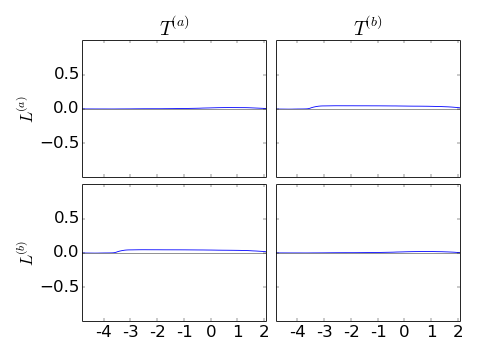}
                \caption{xFDAX}
        \end{subfigure}
        \begin{subfigure}[b]{0.45\textwidth}
                \includegraphics[width=\textwidth,height=40mm]{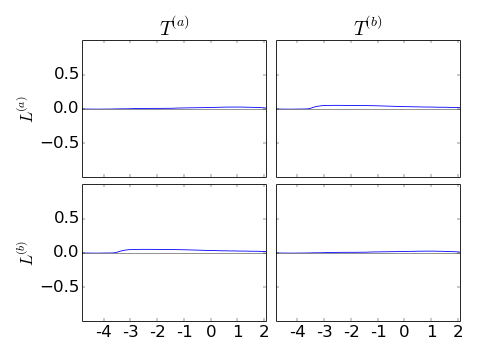}
                \caption{xFGBL}
        \end{subfigure}
\end{figure}

\subsection{Influence of trades on cancel orders}
\label{TC}
\begin{figure}[H]
        \begin{subfigure}[b]{0.45\textwidth}
                \includegraphics[width=\textwidth,height=40mm]{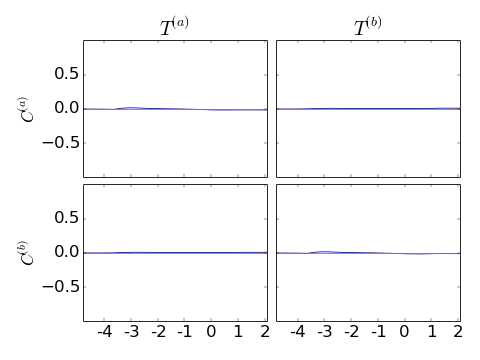}
                \caption{xFDAX}
        \end{subfigure}
        \begin{subfigure}[b]{0.45\textwidth}
                \includegraphics[width=\textwidth,height=40mm]{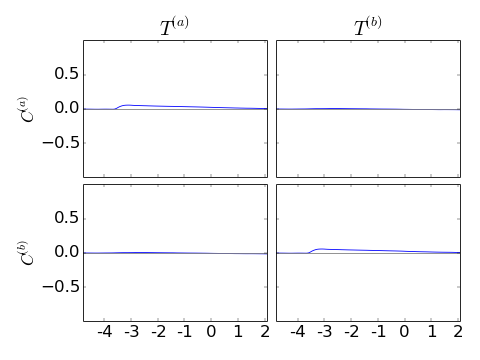}
                \caption{xFGBL}
        \end{subfigure}
\end{figure}

\subsection{Influence of limit orders on trades}
\label{LT}
\begin{figure}[H]
        \begin{subfigure}[b]{0.45\textwidth}
                \includegraphics[width=\textwidth,height=40mm]{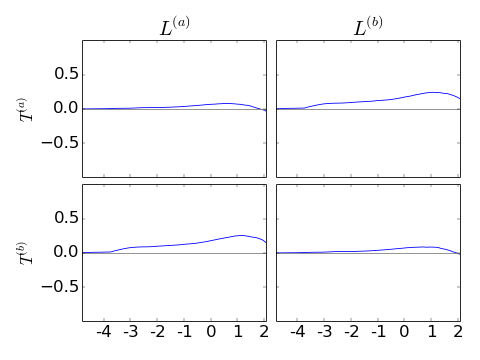}
                \caption{xFDAX}
        \end{subfigure}
        \begin{subfigure}[b]{0.45\textwidth}
                \includegraphics[width=\textwidth,height=40mm]{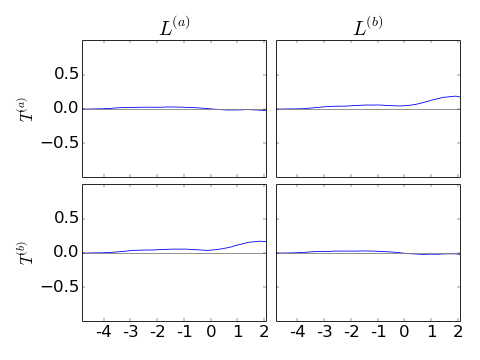}
                \caption{xFGBL}
        \end{subfigure}
\end{figure}

\subsection{Influence of cancel orders on trades}
\label{CT}
\begin{figure}[H]
        \begin{subfigure}[b]{0.45\textwidth}
                \includegraphics[width=\textwidth,height=40mm]{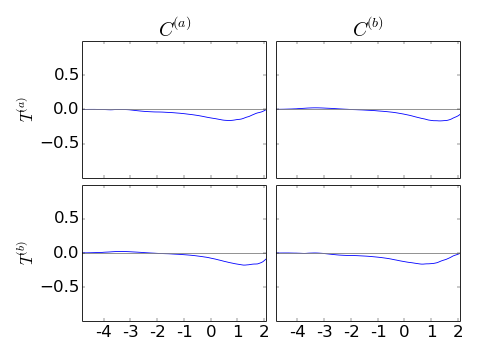}
                \caption{xFDAX}
        \end{subfigure}
        \begin{subfigure}[b]{0.45\textwidth}
                \includegraphics[width=\textwidth,height=40mm]{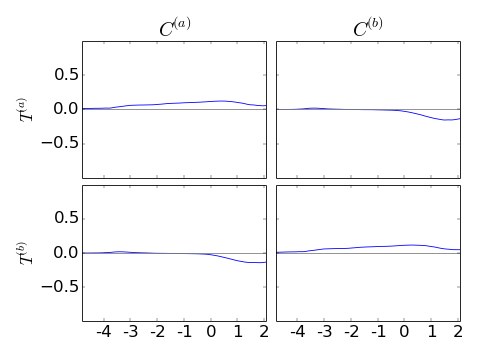}
                \caption{xFGBL}
        \end{subfigure}
\end{figure}

\subsection{Influence of limit orders on cancel orders}
\label{LC}
\begin{figure}[H]
        \begin{subfigure}[b]{0.45\textwidth}
                \includegraphics[width=\textwidth,height=40mm]{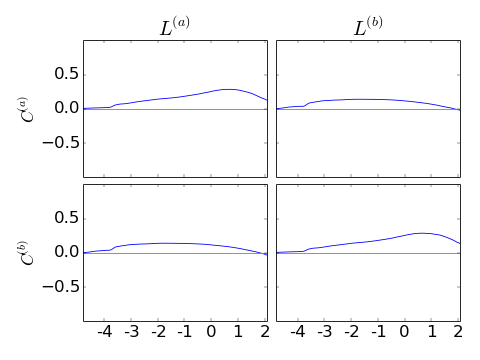}
                \caption{xFDAX}
        \end{subfigure}
        \begin{subfigure}[b]{0.45\textwidth}
                \includegraphics[width=\textwidth,height=40mm]{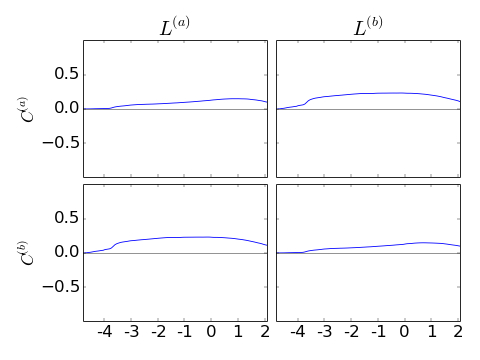}
                \caption{xFGBL}
        \end{subfigure}
\end{figure}

\subsection{Influence of cancel orders on limit orders}
\label{CL}
\begin{figure}[H]
        \begin{subfigure}[b]{0.45\textwidth}
                \includegraphics[width=\textwidth,height=40mm]{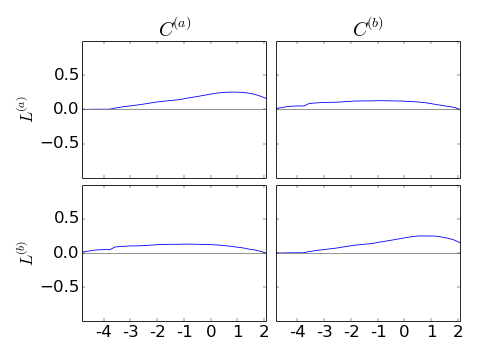}
                \caption{xFDAX}
        \end{subfigure}
        \begin{subfigure}[b]{0.45\textwidth}
                \includegraphics[width=\textwidth,height=40mm]{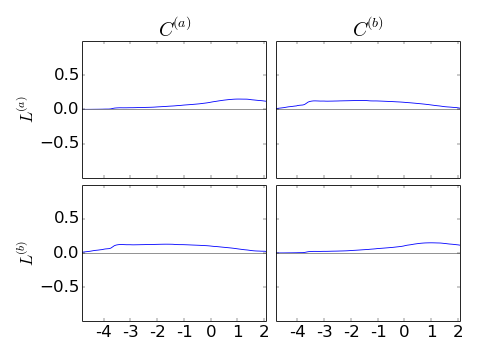}
                \caption{xFGBL}
        \end{subfigure}
\end{figure}

\bibliographystyle{abbrv}
\bibliography{HawkesEstim10}
\end{document}